\documentclass[aps,prl,twocolumn,superscriptaddress,nofootinbib]{revtex4-2}
\usepackage{graphicx}
\usepackage{multirow}
\usepackage{makecell}
\usepackage{color}
\usepackage[dvipsnames]{xcolor}
\usepackage[colorlinks=true,urlcolor=Blue,linkcolor=Blue]{hyperref}
\usepackage[all]{hypcap}
\usepackage{newtxtext,newtxmath}
\usepackage{lipsum}
\usepackage{braket}
\newcommand{\etmax}{\mathrm{e}_{\mathrm{3max}}}
\newcommand{\emax}{\mathrm{e}_{\mathrm{max}}}

\begin{document}

\title{First-Principles Nuclear Modeling for Light Dark Matter Experiments\\ at the Intensity Frontier}

\author{Taylor R.~Gray}
	\email{taylor.gray@chalmers.se}
	\affiliation{Department of Physics and Astronomy, Chalmers University of Technology, 412 96, G\"{o}teborg, Sweden} 
\author{Alberto Scalesi}
	\email{alberto.scalesi@chalmers.se}
    \affiliation{Department of Physics and Astronomy, Chalmers University of Technology, 412 96, G\"{o}teborg, Sweden}

\date{\today}

\begin{abstract}
Accelerator-based experiments at the intensity frontier, in which a high-energy beam impinges on a nuclear target, serve as powerful probes of the light dark matter paradigm. Such experiments require precise modeling of the target nucleus for reliable signal predictions. We present the application of a many-body \textit{ab initio} method to calculate light dark matter mediator production signal rates at electron fixed-target experiments, using chiral effective field theory interactions. Considering both elastic and quasi elastic scattering, we compute cross sections using a Monte Carlo event generator implementation informed by \textit{ab initio} nuclear elastic form factors and spectral functions for three representative nuclei, $^{20}$Ne, $^{34}$Si, and $^{56}$Fe, at varying electron beam energies. We compare our results to a commonly used phenomenological parameterization, finding an increased signal yield by up to two orders of magnitude with our quasi elastic treatment and an agreement for lighter mediators. 

\end{abstract}

\maketitle

\paragraph*{Introduction.}

The challenge of unveiling the fundamental nature of dark matter (DM) is an interdisciplinary effort, drawing on tools and insight from across physics. Decades of null results from direct detection searches targeting GeV-scale Weakly Interacting Massive Particles (WIMPs) have sharpened the case for light DM, a mass range in which thermal freeze-out can also generate the observed relic abundance provided a light mediator connects the dark and visible sectors \cite{Battaglieri:2017aum,Izaguirre:2015yja,Balan:2024cmq}.
Traditional nuclear-recoil direct detection loses sensitivity for sub-GeV DM masses, where the energy transferred to a nucleus falls below typical detector thresholds \cite{Essig:2011nj}, motivating accelerator-based approaches in which the DM or its mediator is produced directly.

Accelerator-based intensity frontier experiments provide a powerful probe of light new physics \cite{Krnjaic:2022ozp,Battaglieri:2017aum}. 
Among these, electron beam fixed-target experiments, in which a high-intensity electron beam impinges on a stationary nuclear target, offer sensitivity to light dark sector particles produced via bremsstrahlung-like radiation off the beam electrons \cite{Bjorken:2009mm}. Several such experiments are currently running or proposed, including NA64 \cite{NA64:2025ddk}, LDMX \cite{LDMX:2025bog, Berger:2026bpq}, DarkShine \cite{DarkSHINE:2024guq}, Lohengrin \cite{Bechtle:2024atq}, HPS experiment \cite{Baltzell:2022rpd}, and past searches including E137 \cite{Batell:2014mga}, employing missing energy, missing momentum, and/or visible decay search strategies, spanning a range of beam energies.

Probing light DM at fixed-target experiments, however, requires precise control over nuclear physics, historically approximated rather than computed from more rigorous nuclear methods.
Current analyses across this experimental program routinely treat the nuclear target through simplified parameterizations \cite{Bjorken:2009mm}, neglecting the many-body nuclear structure known to shape the electron-nucleus response \cite{Benhar:2006wy,Rocco18,Barbieri19}.
This approximation carries real consequences: robust discrimination of a signal excess from Standard Model backgrounds requires both the signal and background rates to be predicted with sufficient precision, and comparing results across experiments and setting exclusion limits requires predictions to be based on consistent and reliable nuclear physics. Moreover, as this class of experiments spans a range of target nuclei, an accurate treatment of nuclear structure offers a valuable tool for experimental design, allowing the choice of target material to be optimized for future searches.

\emph{Ab initio} nuclear many-body methods provide systematically improvable predictions of ground- and excited-state properties of nuclear systems~\cite{Ekstrom23}. Rooted in quantum chromodynamics (QCD) through the chiral effective field theory ($\chi$EFT) interactions they employ~\cite{Machleidt:2016rvv}, they have been proven capable of describing bulk properties of atomic nuclei over a wide region of the nuclear chart~\cite{Stroberg21,Tichai23}. Their reach, however, has long been restricted to (near-)spherical systems; only very recently has the concept of spontaneous symmetry breaking been exploited to extend them to deformed ones~\cite{Hagen22,Frosini22a,Scalesi:2026zrw}, which are characterized by strong long-range correlations and occupy the vast majority of the nuclear chart. While \textit{ab initio} nuclear methods have been applied to DM direct detection~\cite{Gazda:2016mrp,Hu:2021awl,Hoferichter:2019uwa}, their use in modeling the target nuclei of intensity-frontier experiments remains unexplored.

In this Letter we present the first application of a deformed \emph{ab initio} method, the deformed self-consistent Green's function (dSCGF) approach~\cite{Scalesi:2026zrw}, to the prediction of light dark matter (DM) signal rates at fixed-target experiments. Closely related variants of this approach have already been confronted with electron-nucleus scattering observables, yielding good agreement with experimental data for charge densities and form factors~\cite{Duguet17b,Arthuis20} as well as for quasi elastic (QE) cross sections~\cite{Rocco18,Barbieri19}.
We compute the dark mediator production cross section $\sigma$, which sets the expected signal yield $N = \sigma L$ for integrated luminosity $L$. Production proceeds through elastic scattering off the nucleus as a whole and QE scattering off individual nucleons. The elastic contribution is evaluated from the nuclear charge form factor obtained from the dSCGF charge distribution, while the QE contribution is treated in the impulse approximation~\cite{Benhar:2005dj}, with the removal energy and momentum of the struck nucleon distributed according to the dSCGF spectral function~\cite{Rocco:2015cil}.

\paragraph*{Dark matter theory and production.}

While the formalism developed here applies to any light mediator connecting the visible and dark sectors, we focus on a vector mediator, the dark photon $A'$, coupling both to DM and to Standard Model fields. The relevant Lagrangian describing the dark sector interactions is,
\begin{equation}
    -\mathscr{L}_\text{DM} = A'_\mu \left(\epsilon e J^\mu_\text{EM}
    + g_D J_D^\mu\right) ,
    \label{eq:L_A'}
\end{equation}
where $J^\mu_\mathrm{EM} \equiv \sum_f Q_f \bar f \gamma^\mu f$ is the electromagnetic current summed over charged Standard Model fermions $f$ with charge $Q_f$, $J_D^\mu$ is the corresponding DM current, $e$ is the elementary charge, and $\epsilon$ and $g_D$ set the mediator's coupling strength to ordinary matter and DM, respectively. Such a coupling to $J^\mu_\mathrm{EM}$ commonly arises from kinetic mixing between $A'$ and the photon \cite{Holdom:1985ag,Fabbrichesi:2020wbt}, though our analysis does not depend on the details of the underlying UV completion. We consider the sub-GeV mass range from 1 MeV to 1 GeV, where fixed-target experiments are most sensitive. We are agnostic to the mass ratio $m_{A'}/m_{\rm DM}$, the scale of $g_D$, and the specific structure of $J_D^\mu$, since we are only interested in the production rate of on-shell dark photon mediators, which could later decay visibly or invisibly.

\begin{figure}
    \centering
\includegraphics[width=\linewidth]{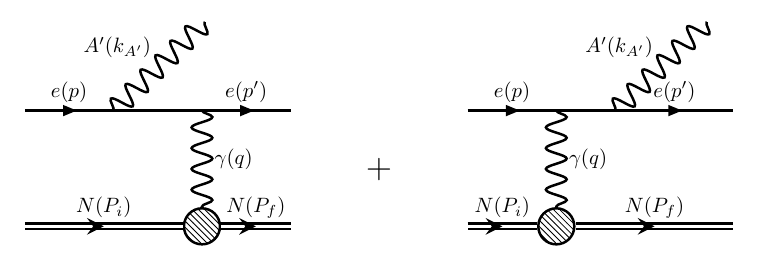}
    \caption{Feynman diagrams for dark photon production through dark bremsstrahlung, where a high-energy electron scatters off a nucleus (elastically or inelastically) and radiates a dark photon from the initial or final state.}
    \label{fig:feynman}
\end{figure}

Due to its coupling to electrons, dark photons can therefore be produced in nuclear fixed-target electron scattering through \textit{dark bremsstrahlung} -- the dark sector analog of ordinary bremsstrahlung. As shown in Fig.~\ref{fig:feynman}, the process proceeds via exchange of a virtual photon carrying four-momentum $q \equiv (\omega, \mathbf{q})$ between the electron and the nucleus, with the dark photon emitted from either the incoming or outgoing electron.
The dark photon could also be emitted off the nucleus through virtual Compton scattering as studied in \cite{Schurmann:2026qmt}, however this diagram contributes sub-dominantly to the total cross section therefore we safely neglect it.

The cross section of $e^- N \to e^- N A'$ is governed by two independent kinematic variables: the three-momentum transfer $|\mathbf{q}|$ and the energy transfer $\omega$ of the virtual photon, with $q^2 = -Q^2 = \omega^2 - |\mathbf{q}|^2 $ and $q^2 < 0$, where $Q$ is a scalar quantity. 
The dark photon mediator mass $m_{A'}$ sets the minimum momentum transfer $Q$ accessible in the bremsstrahlung process, since producing an on-shell mediator of mass $m_{A'}$ requires the target nucleus to absorb at least this much recoil. Light mediators are therefore radiated at small $Q$, where the nucleus responds coherently, while heavy mediators require large $Q$, resolving individual nucleons and pushing the process into the QE regime.
Kinematically, $\omega < |\mathbf{q}|$ (thus $<Q$), so light mediators necessarily also probe small energy transfer $\omega$.
At small $\omega$, thus small mediator masses, the cross section is dominated by elastic scattering, in which the nucleus remains in its ground state and recoils as a whole.
The virtual photon couples coherently to the entire nuclear charge, so the elastic cross section is governed by the nuclear form factor -- the Fourier transform of the ground state charge density.
For larger $\omega$, thus larger mediator masses, the electron instead scatters inelastically with the nucleus. The dominant contribution is where the scattering occurs off the individual nucleons within the nucleus quasi-elastically \cite{Benhar:2006wy}.
The QE peak, broadened by Fermi motion and shifted by the nucleon separation energy, is centered at $\omega \approx Q^2/2m_n$, where $m_n$ is the nucleon mass. 
The cross section also receives contributions from inelastic transitions to discrete nuclear excited states and collective modes, which we neglect in this work.

In standard inclusive electron scattering, $e^-N \to e^-N$, the differential cross section factorizes into a leptonic tensor $L^{\mu\nu}$, contracted with a nuclear tensor $W_{\mu\nu}$ that encodes all the nuclear structure information \cite{Itzykson:1980rh},
\begin{equation}
   d\sigma \propto L^{\mu\nu} W_{\mu\nu} \, ,
    \label{eq:xsec_factorized}
\end{equation}
where the nuclear tensor can be decomposed into two structure functions, $W_1(|\mathbf{q}|,\omega)$ and $W_2(|\mathbf{q}|,\omega)$ \cite{Benhar:2006wy} (see the End Matter for details on these structure functions).
 The dark bremsstrahlung process $e^-N \to e^-N A'$ is governed by the same nuclear tensor, but with a different leptonic tensor~\cite{PhysRevD.8.3109}.
We therefore implement the process exactly at the matrix-element level in \textsc{MadGraph5\_aMC@NLO} \cite{Alwall:2011uj}, a Monte Carlo event generator for automated computation of cross sections. 
The nuclear structure functions are computed from the dSCGF \textit{ab initio} many-body method and implemented via a custom Universal Feynman Output \cite{Darme:2023jdn} model file for input to \textsc{MadGraph}, with details in the End Matter. We compute the dark bremsstrahlung cross section considering both elastic and QE scattering contributions, accounting for Pauli blocking in the latter. Electron screening is neglected; simplified atomic form factor estimates from Ref.~\cite{Bjorken:2009mm} suggest a small effect at low mediator masses, though a treatment consistent with our \textit{ab initio} many-body nuclear structure input is beyond the scope of this work. 

\paragraph*{Ab initio modeling of the target nucleus.}
The dSCGF \emph{ab initio} method employed here is a self-consistent beyond-mean-field approach built on a deformed Hartree--Fock (dHF) reference state~\cite{RingSchuck}. By breaking rotational symmetry, the dHF state efficiently captures the static long-range correlations that dominate in open-shell systems, while dSCGF adds dynamical correlations on top of it~\cite{Soma20b,Soma21}.
The dSCGF expansion must be truncated at a fixed-order, and this is typically done following the so-called Algebraic Diagrammatic Construction (ADC) scheme~\cite{Raimondi18}.
Calculations presented in this work are performed at second order, denoted as dSCGF($2$), and break rotational-symmetry, such that the shape of the computed nuclei is axially-symmetric. A few third-order calculations, denoted as dSCGF($3$), have been performed to estimate the many-body truncation uncertainty.

Two $\chi$EFT interactions are employed, $\Delta$N$^2$LO$_{\rm GO}(394)$~\cite{Jiang20} and N$^3$LO$_{\rm Texas}$~\cite{Hu:2025cjl}. Matrix elements of the two- and three-nucleon (3N) parts of the Hamiltonian are generated with the \texttt{NuHamil} code~\cite{Miyagi23} and expanded in a spherical harmonic-oscillator (sHO) basis, retaining all states up to ${\rm e}_{\max} \equiv \max\{2n+\ell\} = 12$, where $n$ and $\ell$ denote the principal and orbital angular momentum quantum numbers.
The sHO basis is characterized by an oscillator frequency $\hbar\omega$, chosen to optimize the convergence of the observables of interest in all calculations presented in this work.
Three-body operators are included through the rank-reduction procedure of Ref.~\cite{Frosini21}, and the corresponding matrix elements are truncated according to $\etmax \le {\rm e}_{\rm max}^{(1)} + {\rm e}_{\rm max}^{(2)} + {\rm e}_{\rm max}^{(3)}$, the maximum total number of oscillator quanta carried by the three single-particle states.
All calculations employ $\emax = 12$ and $\etmax = 24$ (unless specified otherwise), for which the quantities reported here are converged at the subpercent level, making the associated uncertainty subdominant to the other sources discussed below.

A key advantage of working with a Green's-function-based method is the straightforward access it grants to a broad set of ground- and excited-state observables through the Lehmann representation of the one-body propagator~\cite{Barbieri17}, which encodes the complete single-particle dynamics of the system:
\begin{equation}
\begin{aligned}
g_{\alpha\beta}(\omega) &= \sum_n \frac{\braket{\Psi_0^A | a_\alpha | \Psi_n^{A+1}}\braket{\Psi_n^{A+1} | a_\beta^\dagger | \Psi_0^{A}}}{\omega - \varepsilon_n^+ + i\eta}\\
&+ \sum_k \frac{\braket{\Psi_0^A | a_\beta^\dagger | \Psi_k^{A-1}}\braket{\Psi_k^{A-1} | a_\alpha | \Psi_0^{A}}}{\omega - \varepsilon_k^- - i\eta},
\end{aligned}
\end{equation}
where $\ket{\Psi_0^A}$ is the ground state of the $A$-nucleon system and $\ket{\Psi_n^{A+1}}$, $\ket{\Psi_k^{A-1}}$ are eigenstates of the neighboring systems with one nucleon added or removed; Greek labels run over a complete orthonormal single-particle basis. The poles of the propagator are the one-nucleon separation energies, $\varepsilon_n^+ \equiv E_n^{A+1} - E_0^A$ and $\varepsilon_k^- \equiv E_0^A - E_k^{A-1}$, and the associated spectroscopic amplitudes, $\mathcal{X}_\alpha^n \equiv \braket{\Psi_n^{A+1} | a_\alpha^\dagger | \Psi_0^A}$ and $\mathcal{Y}_\alpha^k \equiv \braket{\Psi_k^{A-1} | a_\alpha | \Psi_0^A}$, quantify the strength carried by each final state in the attachment and knockout channels.

Both quantities required in this work follow directly from these amplitudes.
The hole spectral function entering the QE cross section reads
\begin{equation}
    S(\mathbf{k},E) = \sum_f \big| \big[\, \bra{n_f} \otimes \bra{\Psi_f^{A-1}} \,\big] \ket{\Psi_0^A}  \big|^2 \,
    \delta\!\left(E - E_0^A + E_f^{A-1}\right),
\label{eqn:SF}
\end{equation}
which gives the joint probability of removing a nucleon of momentum $\mathbf{k}$ at removal energy $E$ from the ground state of the nucleus, leaving the residual system in the state $|\Psi_f^{A-1}\rangle$, where the sum is over final states. Here $|n_f\rangle$ denotes the single-particle state of the removed nucleon, carrying momentum $\mathbf{k}$, which couples to the residual $(A-1)$-body eigenstate to form the final state of the process. The one-body density is obtained as
\begin{equation}
\rho_{\alpha\beta} \equiv \braket{\Psi_0^A | a_\beta^\dagger a_\alpha | \Psi_0^A}
= \sum_k (\mathcal{Y}_\beta^k)^*\,\mathcal{Y}_\alpha^k ,
\end{equation}
from which the charge density distribution is generated following
Ref.~\cite{Duguet17b}. Given the deformed setting of the many-body calculations, both the charge density and the spectral function in Eq.~\eqref{eqn:SF} are axially-symmetric.

In order to account for Pauli blocking, necessary for a correct description of the QE response at momentum transfers comparable to the Fermi momentum, we compute the suppression factor $f(|\mathbf{q}\,|)$, whose analytical expression has been taken from Ref.~\cite{Bell:1963ogq,Ballett:2018uuc,Bodek:2021trq} and reads
\begin{equation}
    f(|\mathbf{q}|) =
    \begin{cases}
        \dfrac{3}{4}\dfrac{|\mathbf{q}|}{k_F} - \dfrac{1}{16}\left(\dfrac{|\mathbf{q}|}{k_F}\right)^{3}, & |\mathbf{q}| < 2k_F, \\[2mm]
        1, & |\mathbf{q}| \geq 2k_F.
    \end{cases}
    \label{eq:pauli}
\end{equation}
The Fermi momentum $k_F$ was extracted as the momentum at which the momentum density distribution computed with dSCGF falls to half its value.
The Fermi momenta for each nucleus and $\chi$EFT interaction considered in this work are presented in Table~\ref{tab:kF} for both protons and neutrons.
\begin{table}[t]
\centering
\begin{tabular}{lccc}
\hline\hline
Nucleus & $\chi$EFT interaction & $k_F^p$ [MeV] & $k_F^n$ [MeV] \\
\hline
\multirow{2}{*}{$^{20}$Ne} & N$^3$LO$_{\rm Texas}$           & 173.46 & 175.17 \\
                           & $\Delta$N$^2$LO$_{\rm GO}(394)$ & 166.19 & 168.37 \\
\hline
\multirow{2}{*}{$^{34}$Si} & N$^3$LO$_{\rm Texas}$           & 197.92 & 149.70 \\
                           & $\Delta$N$^2$LO$_{\rm GO}(394)$ & 197.62 & 147.21 \\
\hline
\multirow{2}{*}{$^{56}$Fe} & N$^3$LO$_{\rm Texas}$           & 175.90 & 187.70 \\
                           & $\Delta$N$^2$LO$_{\rm GO}(394)$ & 176.21 & 188.12 \\
\hline\hline
\end{tabular}
\caption{Proton and neutron Fermi momenta used in the Pauli-blocking suppression factor, Eq.~(\ref{eq:pauli}), for each nucleus and $\chi$EFT interaction.}
\label{tab:kF}
\end{table}

\paragraph*{Results.}

\begin{figure*}
    \centering
\includegraphics[width=\linewidth]{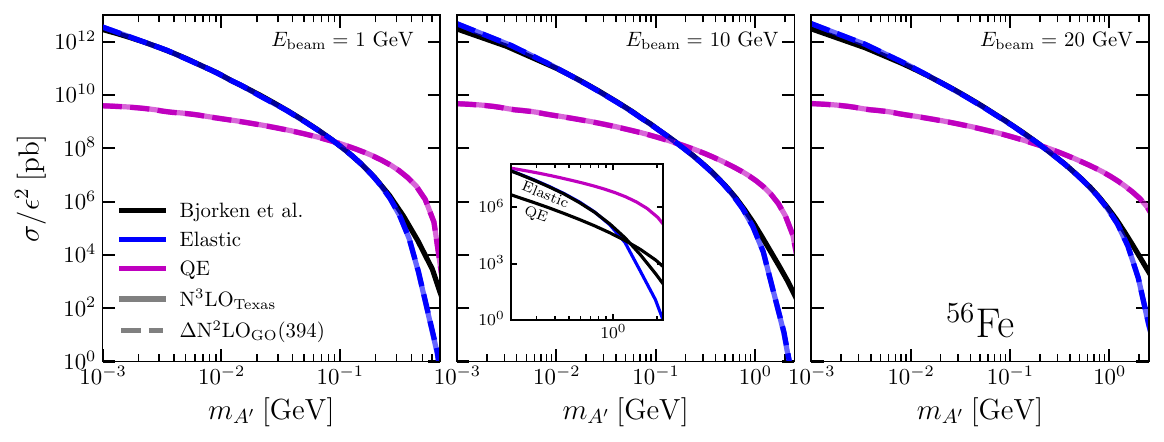}
    \caption{Dark photon mediator production cross section via bremsstrahlung with $^{56}$Fe as a function of the mediator mass $m_{A'}$, for electron beam energies $E_{\rm beam}=1$, $10$, and $20$~GeV (left to right). Elastic (blue) and QE (magenta) contributions are evaluated with the dSCGF nuclear charge form factor and spectral function, respectively, using two $\chi$EFT interactions: N$^3$LO$_{\rm Texas}$ (solid) and $\Delta$N$^2$LO$_{\rm GO} (394)$ (dashed). The analytic parametrization of Ref.~\cite{Bjorken:2009mm}, which combines elastic and inelastic nuclear form factors (Eqs.~A18--A19 of Ref.~\cite{Bjorken:2009mm}), is shown for reference (black). The inset plot in the middle panel compares the elastic and QE components of the Bjorken parametrization. $\sigma$ is divided by $\epsilon^2$, the overall kinetic mixing factor of Eq.~\ref{eq:L_A'}.}
    \label{fig:Fe56}
\end{figure*}

In this Letter we consider three nuclei: $^{20}$Ne, $^{34}$Si, and $^{56}$Fe. They span a broad range of masses, from light to medium-mass systems, as well as of intrinsic deformations: $^{34}$Si is spherical, while $^{20}$Ne and $^{56}$Fe are strongly and weakly prolate, respectively.
Figure~\ref{fig:Fe56} presents the elastic and QE contributions to dark photon mediator production via electron bremsstrahlung off $^{56}$Fe, computed using the Monte Carlo event generator \textsc{MadGraph} with the dSCGF method input.
Below $m_{A'} \sim 0.1$ GeV, with the exact value depending on the nucleus and beam energy, the elastic channel (in blue) dominates, above this mass, the QE channel (in magenta) dominates.

Our results are compared against the widely used analytic parametrization of Ref.~\cite{Bjorken:2009mm}. The elastic form factor is modeled in this parametrization as a smooth, monotonically decreasing function of $Q$ built from a phenomenological charge density, lacking any microscopic nuclear structure.
Moreover, the QE channel is described by an elastic free proton form factor\footnote{Eq. A19 of Ref.~\cite{Bjorken:2009mm} contains a spurious square on the proton dipole form factor term, which should appear to the first power.},
treating the QE scattering as coherent
scattering off a single free proton scaled by $Z$, evaluated at fixed $Q^2$ (lacking any $\omega$ dependence).
This treatment neglects Fermi motion broadening of the QE peak, nuclear binding, inter-nucleon correlations, the neutron contribution, and the effect of Pauli-blocking --- all of which are simultaneously encoded in the spectral function formalism adopted here.

The \textit{ab initio} elastic calculation agrees closely with the elastic parametrization presented in Ref.~\cite{Bjorken:2009mm} for a certain momentum transfer regime.
For $Q$ above $\sim 0.1$ GeV, however, the form factor from the parametrization fails to provide the expected diffractive pattern
\cite{PhysRevLett.38.152,deVries87}. The discrepancies between these treatments of the form factor (as illustrated in the top row of Fig.~\ref{fig:end_matter} in the End Matter) propagate directly into the cross section plotted here. As demonstrated in the inset of Fig.~\ref{fig:Fe56}, for $m_{A'}\gtrsim 1$ GeV the two elastic treatments begin to diverge. 
The elastic channel from Ref.~\cite{Bjorken:2009mm} parametrizes the electron screening, which suppresses the elastic form factor at small momentum transfer. Since it is absent in our formalism, at small mediator masses there is a small discrepancy with our elastic treatment.

In the QE channel, by contrast, the agreement with Ref.~\cite{Bjorken:2009mm} is far weaker. Using the \textit{ab initio} treatment presented in this work, incorporating the nuclear structure functions plotted in the bottom row of Fig.~\ref{fig:end_matter} in the End Matter in place of the analytic parametrization, the  cross section lies systematically above the analytic cross section throughout the regime where the QE channel dominates the mediator production cross section. The enhancement is up to two orders of magnitude from the analytic curve. This discrepancy is not unexpected, since the analytic QE treatment is not informed by nuclear many-body correlations. The strikingly large size of the resulting enhancement signifies that a consistent many-body treatment of the full nuclear spectral function is necessary to obtain a reliable cross section in this regime. This translates directly into a higher projected signal yield and extends the accessible sensitivity to larger DM mediator masses beyond what the phenomenological parametrization of Ref.~\cite{Bjorken:2009mm} would predict. For smaller mediator masses, by contrast, our treatment predicts a reduced QE cross section relative to the analytic parametrization. This reduction is attributed to Pauli blocking, which is absent from the analytic treatment, and our omission of the QE structure functions for $|\mathbf{q}| < 300$~MeV: at such low momentum transfer the impulse approximation is not considered reliable and QE scattering is expected to be suppressed. 
We note that our results (magenta curve) are conservative with respect to the dark mediator signal yield, since including QE scattering beyond the $|\mathbf{q}|$ range considered here and other inelastic channels such as $\Delta$ resonance production would further increase the inelastic cross section.

While $\chi$EFT interaction sensitivity has not been extensively studied, the two interactions used in this work, N$^3$LO$_{\rm Texas}$ and $\Delta$N$^2$LO$_{\rm GO} (394)$ plotted as solid and dashed, respectively, in
Fig.~\ref{fig:Fe56}, show compatible results. The same is found for the other two nuclei considered here.
A few representative calculations have been performed at the dSCGF($3$) level to estimate the impact of truncating the many-body expansion. This turns out to be the largest source of uncertainty among those considered here, amounting to a relative error of $3$--$5\%$ on the total inelastic cross section and $1$--$10\%$ on the elastic one, with the largest deviations occurring at the highest dark photon masses.

\begin{figure}
    \centering
\includegraphics[width=\linewidth]{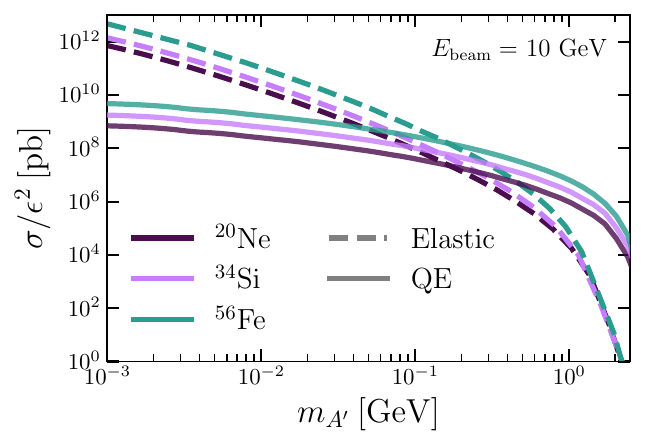}
    \caption{Dark photon mediator production cross section $\sigma$, with $\epsilon^2$ factor divided out, via bremsstrahlung with the nuclei $^{20}$Ne (dark purple), $^{34}$Si (violet), and $^{56}$Fe (teal) as a function of the mediator mass $m_{A'}$. The electron beam energy is fixed to 10 GeV. Elastic (dashed) and QE (solid) contributions are evaluated with the dSCGF nuclear charge form factor and spectral function, respectively, using the N$^3$LO$_{\rm Texas}$ $\chi$EFT interaction.}
    \label{fig:nuclei}
\end{figure}
The cross sections $\sigma/\epsilon^2$ for the nuclei $^{20}$Ne, $^{34}$Si, and $^{56}$Fe at $E_{\rm beam} =
10$~GeV, are given in Fig.~\ref{fig:nuclei}, using the N$^3$LO$_{\rm Texas}$ $\chi$EFT interaction. 
Increasing the nuclear mass enhances both
channels, but with qualitatively different scaling. The elastic cross section is coherently enhanced over the $Z$ protons, so heavier nuclei gain an approximately $Z^2$ boost at fixed $m_{A'}$. The QE channel, governed by the incoherent sum over single nucleon scattering weighted by the spectral function, scales only
linearly with $A$ but retains support out to much larger $m_{A'}$.

\paragraph*{Summary and outlook.}
We demonstrate the imperative of interdisciplinary approaches bridging nuclear many-body theory and particle phenomenology to accurately make new physics predictions. In this work, we focus our analysis on light DM signatures at fixed-target facilities operating at the intensity frontier.
We present cross sections for dark photon mediator production, incorporating state-of-the-art deformed many-body \textit{ab initio} nuclear modeling. We embed our framework for both elastic and QE channels within \textsc{MadGraph}, integrating nuclear structure with Monte Carlo event generation.

The dSCGF many-body \textit{ab initio} method is employed as our nuclear input, which enters the dark mediator production through elastic nuclear form factors and spectral functions, for a representative set of nuclei: $^{20}$Ne, $^{34}$Si, and $^{56}$Fe. The sensitivity to the $\chi$EFT interaction is evaluated by performing calculations with two interactions, N$^3$LO$_{\rm Texas}$ and $\Delta$N$^2$LO$_{\rm GO} (394)$, where we find agreement between both.

While the simple parametrization captures the elastic form factor in restricted kinematic windows, evaluating cross sections across diverse beam energies and target nuclei demands the complete, \textit{ab initio} nuclear framework presented here.
Our QE treatment, which includes the full spectral function formalism of the nuclear structure functions and the effect of Pauli-blocking, notably enhances the sensitivity to light DM for masses $\gtrsim 0.1$ GeV relative to simplified phenomenological nuclear treatments. 
These results highlight the impact of such a framework for providing predictions for light DM mediator searches at intensity-frontier fixed-target experiments, establishing the bridge between systematically improvable first principles nuclear modeling and light DM phenomenology at accelerator-based searches.

Extending the predicted DM mediator event yield to new nuclear targets within our framework requires only a converged many-body calculation of that nucleus, with no additional phenomenological input. This \textit{ab initio} treatment remains applicable well beyond the light- and medium-mass region considered here~\cite{Scalesi:2026zrw}, making it possible to use these predictions directly to compute experimental sensitivities on DM model parameters and guide target material selection in proposed light DM fixed-target searches such as LDMX, Lohengrin, DarkSHINE, and beyond.

\paragraph*{Acknowledgments.}
The authors are grateful to Riccardo Catena and Xavier Roca-Maza for useful comments on the manuscript.
T.G.~has
been funded by the Knut and Alice Wallenberg Foundation, and performed their research
within the “Light Dark Matter” project (Dnr. KAW 2019.0080).
A.S.~acknowledges the Swedish Research Council (Grants No.~2021-04507 and No.~2025-05618).
We acknowledge the National Academic Infrastructure for Supercomputing in Sweden (NAISS), funded by the Swedish Research Council, for providing computational resources.

\bibliography{biblio}

@article{Bjorken:2009mm,
    author = "Bjorken, James D. and Essig, Rouven and Schuster, Philip and Toro, Natalia",
    title = "{New Fixed-Target Experiments to Search for Dark Gauge Forces}",
    eprint = "0906.0580",
    archivePrefix = "arXiv",
    primaryClass = "hep-ph",
    reportNumber = "SLAC-PUB-13650, SU-ITP-09-22",
    doi = "10.1103/PhysRevD.80.075018",
    journal = "Phys. Rev. D",
    volume = "80",
    pages = "075018",
    year = "2009"
}

@article{PhysRevD.8.3109,
  title = {Improved Weizs\"acker-Williams Method and Its Application to Lepton and $W$-Boson Pair Production},
  author = {Kim, Kwang Je and Tsai, Yung-Su},
  journal = {Phys. Rev. D},
  volume = {8},
  issue = {9},
  pages = {3109--3125},
  numpages = {0},
  year = {1973},
  month = {Nov},
  publisher = {American Physical Society},
  doi = {10.1103/PhysRevD.8.3109},
  url = {https://link.aps.org/doi/10.1103/PhysRevD.8.3109}
}

@misc{Krnjaic:2022ozp,
    author = "Krnjaic, G. and others",
    title = "{A Snowmass Whitepaper: Dark Matter Production at Intensity-Frontier Experiments}",
    eprint = "2207.00597",
    archivePrefix = "arXiv",
    primaryClass = "hep-ph",
    reportNumber = "FERMILAB-PUB-22-497-T",
    month = "7",
    year = "2022"
}

@misc{NA64:2025ddk,
    author = "Andreev, Yu. M. and others",
    collaboration = "NA64",
    title = "{Searching for Light Dark Matter and Dark Sectors with the NA64 experiment at the CERN SPS}",
    eprint = "2505.14291",
    archivePrefix = "arXiv",
    primaryClass = "hep-ex",
    month = "5",
    year = "2025"
}

@misc{DarkSHINE:2024guq,
    author = "Chen, Jing and others",
    collaboration = "DarkSHINE",
    title = "{DarkSHINE Baseline Design Report: Physics Prospects and Detector Technologies}",
    eprint = "2411.09345",
    archivePrefix = "arXiv",
    primaryClass = "physics.ins-det",
    month = "11",
    year = "2024"
}

@misc{LDMX:2025bog,
    author = "Akesson, Torsten and others",
    collaboration = "LDMX",
    title = "{LDMX - The Light Dark Matter eXperiment}",
    eprint = "2508.11833",
    archivePrefix = "arXiv",
    primaryClass = "hep-ex",
    year = "2025"
}

@inproceedings{Battaglieri:2017aum,
    author = "Battaglieri, Marco and others",
    title = "{US Cosmic Visions: New Ideas in Dark Matter 2017: Community Report}",
    booktitle = "{U.S. Cosmic Visions: New Ideas in Dark Matter}",
    eprint = "1707.04591",
    archivePrefix = "arXiv",
    primaryClass = "hep-ph",
    reportNumber = "FERMILAB-CONF-17-282-AE-PPD-T",
    month = "7",
    year = "2017"
}

@misc{Schurmann:2026qmt,
    author = {Sch{\"u}rmann, Martin and Dreiner, Herbert K. and Gauld, Rhorry},
    title = "{Theory Calculations for LDMX and LOHENGRIN beyond Coherent Bethe-Heitler Scattering}",
    eprint = "2606.20327",
    archivePrefix = "arXiv",
    primaryClass = "hep-ph",
    reportNumber = "BONN-TH-2026-09, MPP-2026-47",
    month = "6",
    year = "2026"
}

@article{Bechtle:2024atq,
    author = "Bechtle, Philip and others",
    title = "{A proposal for the Lohengrin experiment to search for dark sector particles at the ELSA Accelerator}",
    eprint = "2410.10956",
    archivePrefix = "arXiv",
    primaryClass = "hep-ex",
    doi = "10.1140/epjc/s10052-025-14257-z",
    journal = "Eur. Phys. J. C",
    volume = "85",
    number = "5",
    pages = "600",
    year = "2025"
}

@article{Izaguirre:2015yja,
    author = "Izaguirre, Eder and Krnjaic, Gordan and Schuster, Philip and Toro, Natalia",
    title = "{Analyzing the Discovery Potential for Light Dark Matter}",
    eprint = "1505.00011",
    archivePrefix = "arXiv",
    primaryClass = "hep-ph",
    doi = "10.1103/PhysRevLett.115.251301",
    journal = "Phys. Rev. Lett.",
    volume = "115",
    number = "25",
    pages = "251301",
    year = "2015"
}

@article{Balan:2024cmq,
    author = "Balan, Sowmiya and others",
    title = "{Resonant or asymmetric: the status of sub-GeV dark matter}",
    eprint = "2405.17548",
    archivePrefix = "arXiv",
    primaryClass = "hep-ph",
    reportNumber = "TTP24-015, P3H-24-033",
    doi = "10.1088/1475-7516/2025/01/053",
    journal = "JCAP",
    volume = "01",
    pages = "053",
    year = "2025"
}

@article{Essig:2011nj,
    author = "Essig, Rouven and Mardon, Jeremy and Volansky, Tomer",
    title = "{Direct Detection of Sub-GeV Dark Matter}",
    eprint = "1108.5383",
    archivePrefix = "arXiv",
    primaryClass = "hep-ph",
    reportNumber = "SLAC-PUB-14538",
    doi = "10.1103/PhysRevD.85.076007",
    journal = "Phys. Rev. D",
    volume = "85",
    pages = "076007",
    year = "2012"
}

@article{Alwall:2011uj,
    author = "Alwall, Johan and Herquet, Michel and Maltoni, Fabio and Mattelaer, Olivier and Stelzer, Tim",
    title = "{MadGraph 5 : Going Beyond}",
    eprint = "1106.0522",
    archivePrefix = "arXiv",
    primaryClass = "hep-ph",
    reportNumber = "FERMILAB-PUB-11-448-T",
    doi = "10.1007/JHEP06(2011)128",
    journal = "JHEP",
    volume = "06",
    pages = "128",
    year = "2011"
}

@article{Darme:2023jdn,
    author = "Darm{\'e}, Luc and others",
    title = "{UFO 2.0: the {\textquoteleft}Universal Feynman Output{\textquoteright} format}",
    eprint = "2304.09883",
    archivePrefix = "arXiv",
    primaryClass = "hep-ph",
    reportNumber = "BONN-TH-2023-03, DESY-23-051, FERMILAB-PUB-23-138-T, KA-TP-06-2023,
  MCNET-23-06, P3H-23-023, TIF-UNIMI-2023-11",
    doi = "10.1140/epjc/s10052-023-11780-9",
    journal = "Eur. Phys. J. C",
    volume = "83",
    number = "7",
    pages = "631",
    year = "2023"
}

@misc{Arthuis20,
      author         = "Arthuis, P. and Barbieri, C. and Vorabbi, M. and Finelli,
                        P.",
      title          = "{Ab initio computation of charge densities for Sn and Xe
                        isotopes}",
      year           = "2020",
      eprint         = "2002.02214",
      archivePrefix  = "arXiv",
      primaryClass   = "nucl-th"
}

@incollection{Barbieri17,
  author={Barbieri, C. and Carbone, A.},
  title={Self-Consistent {G}reen's Function Approaches},
  booktitle={An Advanced Course in Computational Nuclear Physics},
  publisher={Springer, Cham},
  editor={Hjorth-Jensen, M. and Lombardo, M. and van Kolck, U.},
  series={Lecture Notes in Physics},
  volume={936},
  year={2017}
}

@article{Barbieri19,
  title = {Lepton scattering from $^{40}\mathrm{Ar}$ and $^{48}\mathrm{Ti}$ in the quasielastic peak region},
  author = {Barbieri, C. and Rocco, N. and Som\`a, V.},
  journal = {Phys. Rev. C},
  volume = {100},
  issue = {6},
  pages = {062501},
  numpages = {6},
  year = {2019},
  month = {12},
  publisher = {American Physical Society},
  doi = {10.1103/PhysRevC.100.062501},
  url = {https://link.aps.org/doi/10.1103/PhysRevC.100.062501}
}

@article{Benhar:2006wy,
    author = "Benhar, Omar and day, Donal and Sick, Ingo",
    title = "{Inclusive quasi-elastic electron-nucleus scattering}",
    eprint = "nucl-ex/0603029",
    archivePrefix = "arXiv",
    doi = "10.1103/RevModPhys.80.189",
    journal = "Rev. Mod. Phys.",
    volume = "80",
    pages = "189--224",
    year = "2008"
}

@article{DeForest:1966ycn,
    author = "De Forest, Jr., T. and Walecka, J. D.",
    title = "{Electron scattering and nuclear structure}",
    doi = "10.1080/00018736600101254",
    journal = "Adv. Phys.",
    volume = "15",
    pages = "1--109",
    year = "1966"
}

@article{Holdom:1985ag,
    author = "Holdom, Bob",
    title = "{Two U(1)'s and Epsilon Charge Shifts}",
    reportNumber = "UTPT-85-30",
    doi = "10.1016/0370-2693(86)91377-8",
    journal = "Phys. Lett. B",
    volume = "166",
    pages = "196--198",
    year = "1986"
}

@book{Itzykson:1980rh,
    author = "Itzykson, C. and Zuber, J. B.",
    title = "{Quantum Field Theory}",
    isbn = "978-0-486-44568-7",
    publisher = "McGraw-Hill",
    address = "New York",
    series = "International Series In Pure and Applied Physics",
    year = "1980"
}

@misc{Fabbrichesi:2020wbt,
    author = "Fabbrichesi, Marco and Gabrielli, Emidio and Lanfranchi, Gaia",
    title = "{The Dark Photon}",
    eprint = "2005.01515",
    archivePrefix = "arXiv",
    primaryClass = "hep-ph",
    doi = "10.1007/978-3-030-62519-1",
    month = "5",
    year = "2020"
}

@article{deVries87,
      author         = "De Vries, H. and De Jager, C. W. and De Vries, C.",
      title          = "{Nuclear charge and magnetization density distribution
                        parameters from elastic electron scattering}",
      journal        = "Atom. Data Nucl. Data Tabl.",
      volume         = "36",
      year           = "1987",
      pages          = "495-536",
      doi            = "10.1016/0092-640X(87)90013-1",
      SLACcitation   = "%%CITATION = ADNDA,36,495;%%"
}

@article{Duguet17b,
  title = {Ab initio calculation of the potential bubble nucleus $^{34}\mathrm{Si}$},
  author = {Duguet, T. and Som\`a, V. and Lecluse, S. and Barbieri, C. and Navr\'atil, P.},
  journal = {Phys. Rev. C},
  volume = {95},
  issue = {3},
  pages = {034319},
  numpages = {17},
  year = {2017},
  month = {Mar},
  publisher = {American Physical Society},
  doi = {10.1103/PhysRevC.95.034319},
  url = {https://link.aps.org/doi/10.1103/PhysRevC.95.034319}
}

@article{Ekstrom23,
  author={Ekstr\"om, A.  and Forss\'en, C.  and Hagen, G.  and Jansen, G. R.  and Jiang, W.  and Papenbrock, T. },
  title={What is ab initio in nuclear theory?},
  journal={Frontiers in Physics},
  volume={11},
  year={2023},
  url={https://www.frontiersin.org/journals/physics/articles/10.3389/fphy.2023.1129094},
  doi={10.3389/fphy.2023.1129094}
}

@article{Frosini21,
    author = "Frosini, M. and Duguet, T. and Bally, B. and Beaujeault-Taudi\`ere, Y. and Ebran, J. -P. and Som\`a, V.",
    title = "{In-medium $k$-body reduction of $n$-body operators: A flexible symmetry-conserving approach based on the sole one-body density matrix}",
    eprint = "2102.10120",
    archivePrefix = "arXiv",
    primaryClass = "nucl-th",
    doi = "10.1140/epja/s10050-021-00458-z",
    journal = "Eur. Phys. J. A",
    volume = "57",
    number = "4",
    pages = "151",
    year = "2021"
}

@article{Frosini22a,
    author = "Frosini, M. and Duguet, T. and Ebran, J.-P. and Som\`a, V.",
    title = "{Multi-reference many-body perturbation theory for nuclei: I. Novel PGCM-PT formalism}",
    eprint = "2110.15737",
    archivePrefix = "arXiv",
    primaryClass = "nucl-th",
    doi = "10.1140/epja/s10050-022-00692-z",
    journal = "Eur. Phys. J. A",
    volume = "58",
    number = "4",
    pages = "62",
    year = "2022"
}

@article{Hagen22,
  title = {Angular-momentum projection in coupled-cluster theory: Structure of $^{34}\mathrm{Mg}$},
  author = {Hagen, G. and Novario, S. J. and Sun, Z. H. and Papenbrock, T. and Jansen, G. R. and Lietz, J. G. and Duguet, T. and Tichai, A.},
  journal = {Phys. Rev. C},
  volume = {105},
  issue = {6},
  pages = {064311},
  numpages = {23},
  year = {2022},
  month = {6},
  publisher = {American Physical Society},
  doi = {10.1103/PhysRevC.105.064311}
}

@article{Hu:2021awl,
    author = {Hu, B. S. and Padua-Arg{\"u}elles, J. and Leutheusser, S. and Miyagi, T. and Stroberg, S. R. and Holt, J. D.},
    title = "{Ab~Initio Structure Factors for Spin-Dependent Dark Matter Direct Detection}",
    eprint = "2109.00193",
    archivePrefix = "arXiv",
    primaryClass = "nucl-th",
    doi = "10.1103/PhysRevLett.128.072502",
    journal = "Phys. Rev. Lett.",
    volume = "128",
    number = "7",
    pages = "072502",
    year = "2022"
}

@article{Jiang20,
  title = {Accurate bulk properties of nuclei from $A=2$ to $\ensuremath{\infty}$ from potentials with $\mathrm{\ensuremath{\Delta}}$ isobars},
  author = {Jiang, W. G. and Ekstr\"om, A. and Forss\'en, C. and Hagen, G. and Jansen, G. R. and Papenbrock, T.},
  journal = {Phys. Rev. C},
  volume = {102},
  issue = {5},
  pages = {054301},
  numpages = {8},
  year = {2020},
  month = {11},
  publisher = {American Physical Society},
  doi = {10.1103/PhysRevC.102.054301},
  url = {https://link.aps.org/doi/10.1103/PhysRevC.102.054301}
}

@article{Miyagi23,
  title={NuHamil : A numerical code to generate nuclear two- and three-body matrix elements from chiral effective field theory},
  author={T. Miyagi},
  journal={Eur. Phys. J. A},
  year={2023},
  volume={59},
  doi = "10.1140/epja/s10050-023-01039-y",
  url={https://api.semanticscholar.org/CorpusID:256900656}
}

@article{Raimondi18,
  title = {Algebraic diagrammatic construction formalism with three-body interactions},
  author = {Raimondi, F. and Barbieri, C.},
  journal = {Phys. Rev. C},
  volume = {97},
  issue = {5},
  pages = {054308},
  numpages = {31},
  year = {2018},
  month = {5},
  publisher = {American Physical Society},
  doi = {10.1103/PhysRevC.97.054308},
  url = {https://link.aps.org/doi/10.1103/PhysRevC.97.054308}
}

@book{RingSchuck,
   author = "P. Ring and P. Schuck",
   year = "1980",
   title = "The Nuclear Many-Body Problem",
   publisher = "Springer-Verlag",
   address = "New-York"
}

@article{Rocco:2015cil,
    author = "Rocco, Noemi and Lovato, Alessandro and Benhar, Omar",
    title = "{Unified description of electron-nucleus scattering within the spectral function formalism}",
    eprint = "1512.07426",
    archivePrefix = "arXiv",
    primaryClass = "nucl-th",
    doi = "10.1103/PhysRevLett.116.192501",
    journal = "Phys. Rev. Lett.",
    volume = "116",
    number = "19",
    pages = "192501",
    year = "2016"
}

@article{Rocco:2020jlx,
    author = "Rocco, Noemi",
    title = "{Ab initio Calculations of Lepton-Nucleus Scattering}",
    reportNumber = "FERMILAB-PUB-20-309-T",
    doi = "10.3389/fphy.2020.00116",
    journal = "Front. in Phys.",
    volume = "8",
    pages = "116",
    year = "2020"
}

@article{DeForest:1984qe,
    author = "De Forest, T.",
    title = "{{The relativistic Coulomb sum rule for electron scattering in the independent particle model}}",
    doi = "10.1016/0375-9474(84)90607-9",
    journal = "Nucl. Phys. A",
    volume = "414",
    pages = "347--358",
    year = "1984"
}

@article{Perdrisat:2006hj,
    author = "Perdrisat, C. F. and Punjabi, V. and Vanderhaeghen, M.",
    title = "{Nucleon Electromagnetic Form Factors}",
    eprint = "hep-ph/0612014",
    archivePrefix = "arXiv",
    reportNumber = "WM-06-115, JLAB-THY-06-595",
    doi = "10.1016/j.ppnp.2007.05.001",
    journal = "Prog. Part. Nucl. Phys.",
    volume = "59",
    pages = "694--764",
    year = "2007"
}

@article{Ballett:2018uuc,
    author = "Ballett, Peter and Hostert, Matheus and Pascoli, Silvia and Perez-Gonzalez, Yuber F. and Tabrizi, Zahra and Zukanovich Funchal, Renata",
    title = "{Neutrino Trident Scattering at Near Detectors}",
    eprint = "1807.10973",
    archivePrefix = "arXiv",
    primaryClass = "hep-ph",
    reportNumber = "IPPP/18/64",
    doi = "10.1007/JHEP01(2019)119",
    journal = "JHEP",
    volume = "01",
    pages = "119",
    year = "2019"
}

@article{Benhar:2005dj,
    author = "Benhar, Omar and Farina, Nicola and Nakamura, Hiroki and Sakuda, Makoto and Seki, Ryoichi",
    title = "{Electron- and neutrino-nucleus scattering in the impulse approximation regime}",
    eprint = "hep-ph/0506116",
    archivePrefix = "arXiv",
    doi = "10.1103/PhysRevD.72.053005",
    journal = "Phys. Rev. D",
    volume = "72",
    pages = "053005",
    year = "2005"
}

@article{Carlson:2001mp,
    author = "Carlson, J. and Jourdan, J. and Schiavilla, R. and Sick, I.",
    title = "{Longitudinal and transverse quasielastic response functions of light nuclei}",
    eprint = "nucl-th/0106047",
    archivePrefix = "arXiv",
    reportNumber = "LA-UR-01-3235, JLAB-THY-01-19",
    doi = "10.1103/PhysRevC.65.024002",
    journal = "Phys. Rev. C",
    volume = "65",
    pages = "024002",
    year = "2002"
}

@misc{Baltzell:2022rpd,
    author = "Baltzell, Nathan and others",
    title = "{The Heavy Photon Search Experiment}",
    eprint = "2203.08324",
    archivePrefix = "arXiv",
    primaryClass = "hep-ex",
    month = "3",
    year = "2022"
}

@article{Batell:2014mga,
    author = "Batell, Brian and Essig, Rouven and Surujon, Ze'ev",
    title = "{Strong Constraints on Sub-GeV Dark Sectors from SLAC Beam Dump E137}",
    eprint = "1406.2698",
    archivePrefix = "arXiv",
    primaryClass = "hep-ph",
    doi = "10.1103/PhysRevLett.113.171802",
    journal = "Phys. Rev. Lett.",
    volume = "113",
    number = "17",
    pages = "171802",
    year = "2014"
}

@misc{Berger:2026bpq,
    author = "Berger, Adam and Catena, Riccardo and Conrad, Jan and Gray, Taylor R.",
    title = "{Light Dark Matter Discovery Potential and Model Selection at LDMX}",
    eprint = "2607.24524",
    archivePrefix = "arXiv",
    primaryClass = "hep-ph",
    month = "7",
    year = "2026"
}

@misc{Bodek:2021trq,
    author = "Bodek, Arie",
    title = "{Pauli Blocking for a Relativistic Fermi Gas in Quasielastic Lepton Nucleus Scattering}",
    eprint = "2111.03631",
    archivePrefix = "arXiv",
    primaryClass = "nucl-th",
    month = "11",
    year = "2021"
}

@article{Gazda:2016mrp,
    author = "Gazda, Daniel and Catena, Riccardo and Forss{\'e}n, Christian",
    title = "{Ab initio nuclear response functions for dark matter searches}",
    eprint = "1612.09165",
    archivePrefix = "arXiv",
    primaryClass = "hep-ph",
    doi = "10.1103/PhysRevD.95.103011",
    journal = "Phys. Rev. D",
    volume = "95",
    number = "10",
    pages = "103011",
    year = "2017"
}

@article{Galster:1971kv,
    author = "Galster, S. and Klein, H. and Moritz, J. and Schmidt, K. H. and Wegener, D. and Bleckwenn, J.",
    title = "{Elastic electron-deuteron scattering and the electric neutron form factor at four-momentum transfers 5fm$^{-2} < q^2 < 14$fm$^{-2}$}",
    reportNumber = "DESY-71-7",
    doi = "10.1016/0550-3213(71)90068-X",
    journal = "Nucl. Phys. B",
    volume = "32",
    pages = "221--237",
    year = "1971"
}

@article{Bell:1963ogq,
    author = "Bell, J. S. and Veltman, M. J. G.",
    title = "{Intermediate boson production by neutrinos}",
    doi = "10.1016/S0375-9601(63)80045-6",
    journal = "Phys. Lett.",
    volume = "5",
    pages = "94--96",
    year = "1963"
}

@article{Rocco18,
  title = {Inclusive electron-nucleus cross section within the self-consistent Green's function approach},
  author = {Rocco, N. and Barbieri, C.},
  journal = {Phys. Rev. C},
  volume = {98},
  issue = {2},
  pages = {025501},
  numpages = {11},
  year = {2018},
  month = {8},
  publisher = {American Physical Society},
  doi = {10.1103/PhysRevC.98.025501},
  url = {https://link.aps.org/doi/10.1103/PhysRevC.98.025501}
}

@article{Machleidt:2016rvv,
    author = "Machleidt, R. and Sammarruca, F.",
    title = "{Chiral EFT based nuclear forces: Achievements and challenges}",
    eprint = "1608.05978",
    archivePrefix = "arXiv",
    primaryClass = "nucl-th",
    doi = "10.1088/0031-8949/91/8/083007",
    journal = "Phys. Scripta",
    volume = "91",
    number = "8",
    pages = "083007",
    year = "2016"
}

@article{Soma20b,
    author = "Som\`a, V.",
    title = "{Self-consistent Green's function theory for atomic nuclei}",
    eprint = "2003.11321",
    archivePrefix = "arXiv",
    primaryClass = "nucl-th",
    doi = "10.3389/fphy.2020.00340",
    journal = "Front. in Phys.",
    volume = "8",
    pages = "340",
    year = "2020"
}

@article{Soma21,
    author = "Som\`a, V. and Barbieri, C. and Duguet, T. and Navr\'atil, P.",
    title = "{Moving away from singly-magic nuclei with Gorkov Green's function theory}",
    eprint = "2009.01829",
    archivePrefix = "arXiv",
    primaryClass = "nucl-th",
    doi = "10.1140/epja/s10050-021-00437-4",
    journal = "Eur. Phys. J. A",
    volume = "57",
    number = "4",
    pages = "135",
    year = "2021"
}

@article{Stroberg21,
    author = {Stroberg, S. R. and Holt, J. D. and Schwenk, A. and Simonis, J.},
    doi = {10.1103/PhysRevLett.126.022501},
    issue = {2},
    journal = {Phys. Rev. Lett.},
    month = {1},
    numpages = {6},
    pages = {022501},
    publisher = {American Physical Society},
    title = {Ab Initio Limits of Atomic Nuclei},
    url = {https://link.aps.org/doi/10.1103/PhysRevLett.126.022501},
    volume = {126},
    year = {2021}
}

@article{Tichai23,
    author = "Tichai, A. and Demol, P. and Duguet, T.",
    title = "{Towards heavy-mass ab initio nuclear structure: Open-shell Ca, Ni and Sn isotopes from Bogoliubov coupled-cluster theory}",
    doi = "10.1016/j.physletb.2024.138571",
    journal = "Phys. Lett. B",
    volume = "851",
    pages = "138571",
    year = "2024"
}

@misc{Hu:2025cjl,
    author = {Hu, B. S. and Ekstr{\"o}m, A. and Forss{\'e}n, C. and Hagen, G. and Jiang, W. G. and Miyagi, T. and Papenbrock, T.},
    title = "{The neutron dripline in calcium isotopes from a chiral interaction}",
    eprint = "2512.11723",
    archivePrefix = "arXiv",
    primaryClass = "nucl-th",
    month = "12",
    year = "2025"
}

@article{PhysRevLett.38.152,
  title = {High-Momentum-Transfer Electron Scattering from $^{208}\mathrm{Pb}$},
  author = {Frois, B. and Bellicard, J. B. and Cavedon, J. M. and Huet, M. and Leconte, P. and Ludeau, P. and Nakada, A. and H\^o, Phan Zuan and Sick, I.},
  journal = {Phys. Rev. Lett.},
  volume = {38},
  issue = {4},
  pages = {152--155},
  numpages = {0},
  year = {1977},
  month = {Jan},
  publisher = {American Physical Society},
  doi = {10.1103/PhysRevLett.38.152},
  url = {https://link.aps.org/doi/10.1103/PhysRevLett.38.152}
}

@article{Hoferichter:2019uwa,
    author = "Hoferichter, Martin and Klos, Philipp and Men{\'e}ndez, Javier and Schwenk, Achim",
    title = "{Dark-matter-nucleus scattering in chiral effective field theory}",
    eprint = "1903.11075",
    archivePrefix = "arXiv",
    primaryClass = "hep-ph",
    doi = "10.22323/1.317.0095",
    journal = "PoS",
    volume = "CD2018",
    pages = "095",
    year = "2019"
}

@misc{Scalesi:2026zrw,
    author = "Scalesi, A. and Duguet, T. and Som{\`a}, V.",
    title = "{Deformed self-consistent Green's function method for atomic nuclei at second and third order in the algebraic diagrammatic construction}",
    eprint = "2608.23700",
    archivePrefix = "arXiv",
    primaryClass = "nucl-th",
    month = "8",
    year = "2026"
}

\clearpage
\section*{End Matter}

\begin{figure}[h!]
    \centering
    \includegraphics[width=\linewidth]{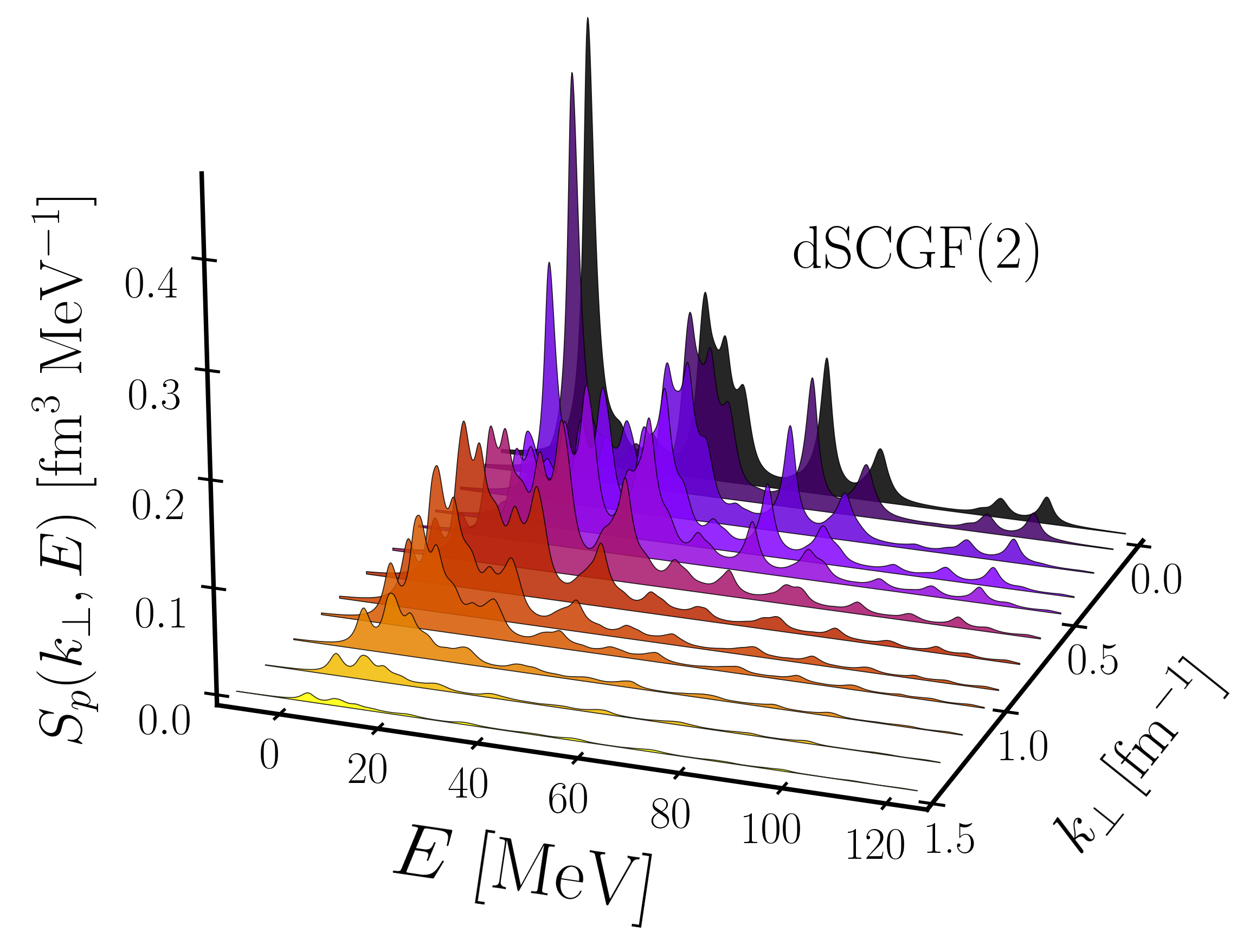}
    \includegraphics[width=\linewidth]{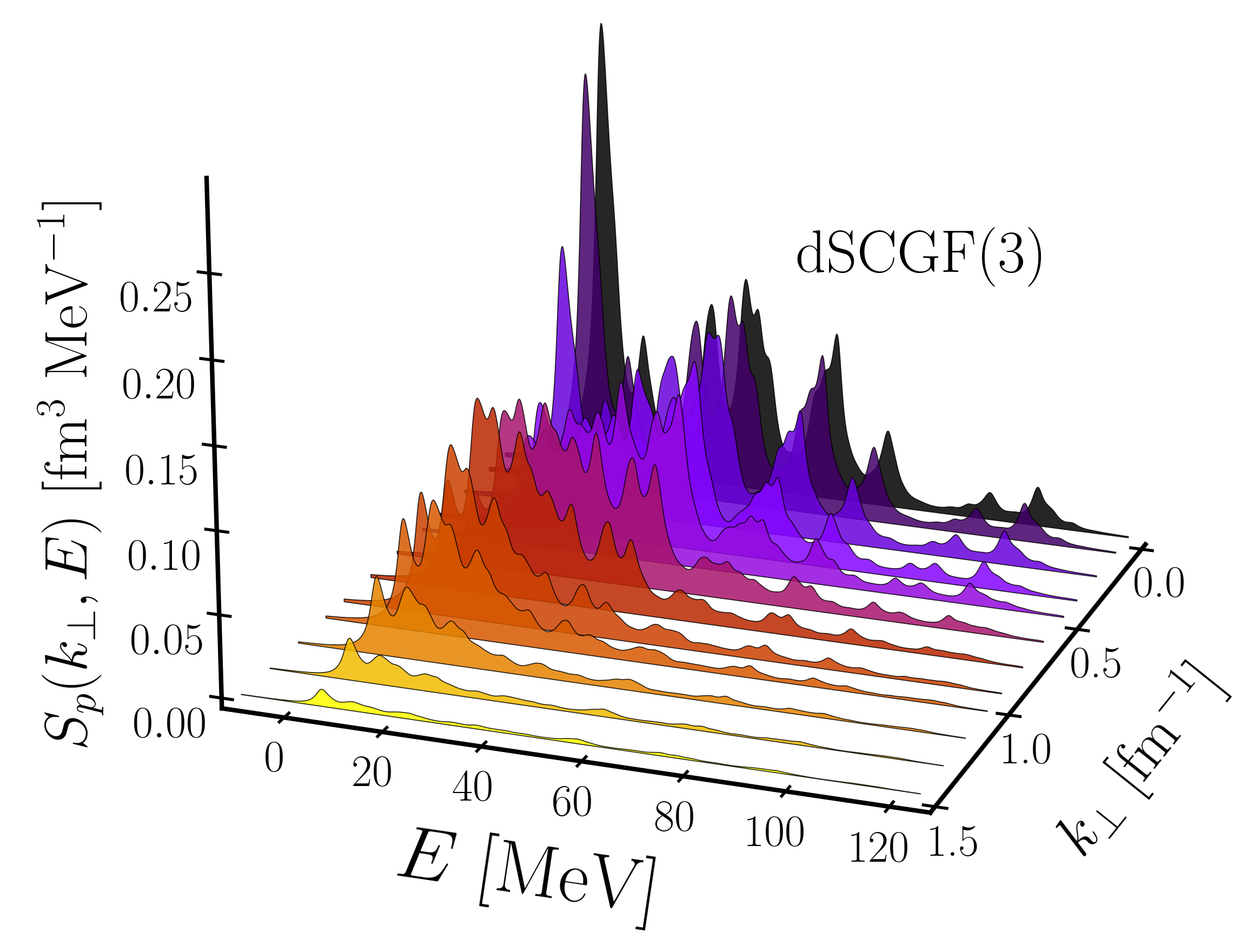}
    \caption{Proton spectral function of $^{56}$Fe at $k_z = 0$~fm$^{-1}$, computed in the dSCGF(2) and dSCGF(3) approximations.
    Peaks are smeared with a Lorentzian of width $\Gamma = 1.5$~MeV for display purposes.
    Calculations employ the $\Delta$N$^2$LO$_{\rm GO}(394)$ interaction with $\emax = 6$, $\etmax = 18$ and an oscillator frequency $\hbar\omega = 18$~MeV.}
    \label{fig:SF}
\end{figure}

\paragraph{Nuclear structure functions for quasi elastic scattering.}
We make use of the impulse approximation, in which QE scattering reduces to a sum of scatterings off individual bound nucleons within the nucleus \cite{Benhar:2006wy}. The nuclear spectral function in Eq.~\eqref{eqn:SF} directly enters the QE nuclear structure functions $W_1$ and $W_2$, through \cite{DeForest:1984qe,Benhar:2006wy,Rocco:2020jlx},
\begin{equation}
\begin{aligned}
W_1(|\mathbf{q}|,\omega) &= \int d^3k\,dE \Bigg[ ZS_p(\mathbf{k},E) \frac{m_n}{E_k} \\ & \Big[ w_1^p(|\mathbf{q}|,\tilde{\omega}) + w_2^p(|\mathbf{q}|,\tilde{\omega})\frac{|\mathbf{k}\times\mathbf{q}|^2}{2m_n^2|\mathbf{q}|^2}\Big] + (n)
  \Bigg],
\end{aligned}
\end{equation}
\begin{equation}
\begin{aligned}
&W_2(|\mathbf{q}|,\omega) = \int d^3k\,dE \Bigg[  ZS_p(\mathbf{k},E) \frac{m_n}{E_k} \\ & \Big[ w_1^p(|\mathbf{q}|,\tilde{\omega}) \frac{q^2}{|\mathbf{q}|^2} \left( \frac{q^2}{\tilde{q}^2}-1 \right) + \frac{w_2^p(|\mathbf{q}|,\tilde{\omega})}{m_n^2}\Bigg( \frac{q^4}{|\mathbf{q}|^4} \\ &  \left( E_k - \frac{\tilde{\omega} ( E_k \tilde{\omega}-\mathbf{k}\cdot \mathbf{q})}{\tilde{q}^2} \right)^2  -\frac{q^2 |\mathbf{k} \times \mathbf{q}|^2}{2|\mathbf{q}|^4}\Bigg) \Big] + (n) \Bigg],
\end{aligned}
\label{eq:W1W2}
\end{equation}
where $\mathbf k$ and $E_k$ are the initial momentum and energy of the struck nucleon, $(n)$ denotes the analogous neutron term weighted by $A-Z$ instead of $Z$, and we take the $w_1^N$ and $w_2^N$ structure functions using the dipole and Galster parameterizations \cite{Galster:1971kv, Perdrisat:2006hj}.
The proton component of the spectral function, $S_{p}(\mathbf{k},E)$, defined in Eq.~\ref{eqn:SF}, is represented in Fig.~\ref{fig:SF} for dSCGF(2) and dSCGF(3).
Increasing the order of the many-body truncation enhances the fragmentation of the peaks, in line with what has been observed in previous studies~\cite{Soma20b}.

We restrict the QE calculation to $300\ \text{MeV} \lesssim |\mathbf q\,| \lesssim 800\ \text{MeV}$, the window in which the impulse approximation is expected to hold \cite{Carlson:2001mp,Rocco:2020jlx}: large enough that the probe resolves individual nucleons, and small enough to remain below the onset of $\Delta$ and $\pi$ production channels \cite{Benhar:2006wy}.
The nucleon is bound, so only part of the energy transferred to the nucleus reaches the struck nucleon; this is accounted for by the shifted four-momentum $\tilde q \equiv (\tilde\omega,\mathbf{q}\,)$, with $\tilde\omega \equiv \omega - E + m_n - E_k$ and the nucleon mass $m_n$ \cite{Benhar:2006wy}. 

\begin{figure*}
    \centering
    \includegraphics[width=0.75\linewidth]{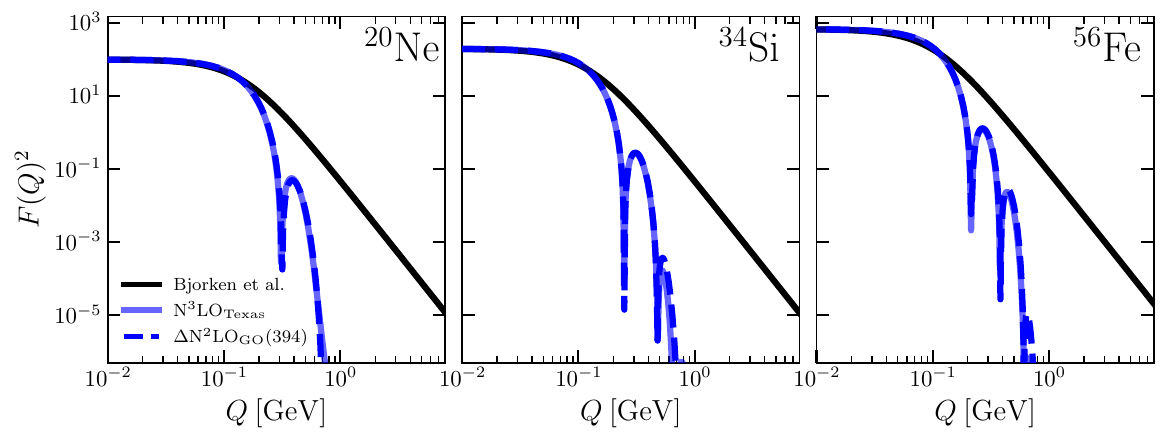}
    \includegraphics[width=0.75\linewidth]{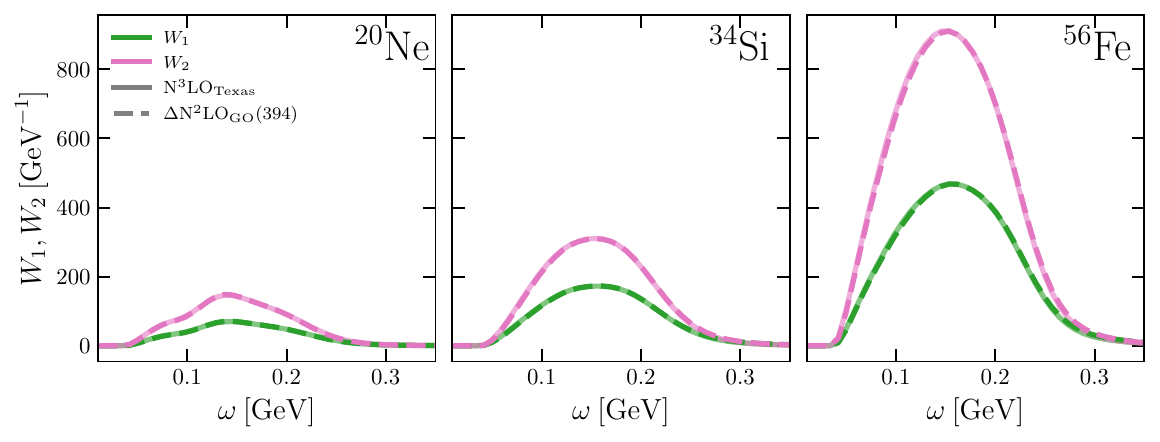}
    \caption{Elastic form factors and QE structure functions for $^{20}$Ne, $^{34}$Si, and $^{56}$Fe. \textit{Top row:} elastic form factor squared, $F(Q)^2$, as a function of momentum transfer $Q\equiv \sqrt{-q^2}$, comparing the phenomenological parametrization from Ref.~\cite{Bjorken:2009mm} (black) with two $\chi$EFT interactions, N$^3$LO$_{\rm Texas}$ (blue, solid) and $\Delta$N$^2$LO$_{\rm GO}(394)$ (dark blue, dashed). \textit{Bottom row:} QE structure functions $W_1$ (green) and $W_2$ (pink) at fixed $\mathbf{q}=500$ MeV, as functions of energy transfer $\omega$, computed from the N$^3$LO$_{\rm Texas}$ (solid) and $\Delta$N$^2$LO$_{\rm GO}(394)$ (dashed) interactions via the dSCGF spectral function.}
    \label{fig:end_matter}
\end{figure*}

The QE nuclear structure functions $W_1$ and $W_2$ for each nucleus and $\chi$EFT interaction are plotted as a function of energy transfer in the bottom row of Fig.~\ref{fig:end_matter}, for fixed $|\mathbf{q}|=500$ MeV. Notice that, similarly to the elastic form factor, both $\chi$EFT interactions coincide with each other. The nuclear structure functions grow with nucleon number, reflecting the increasing number of protons and neutrons available to absorb the momentum transfer.

\paragraph{Elastic nuclear form factor.}

The elastic nuclear form factor is computed from the charge density $\rho_{\rm ch}(r)$ \cite{DeForest:1966ycn, Duguet17b},
\begin{equation}
    F(Q^2) = \int d^3r\, \rho_{\rm ch}(r)\, e^{i\mathbf q \cdot \mathbf r},
    \label{eq:F(q)}
\end{equation}
normalized such that $F(0)=Z$, and since elastic scattering occurs at $\omega=0$ in the nuclear rest frame, the invariant four-momentum transfer $q^2$ reduces to $-|\mathbf q\,|^2$, thus evaluating the Fourier transform with $\mathbf{q}$ yields the frame-independent $F(Q^2)$.
The squared elastic nuclear form factors are plotted in the top row of Fig.~\ref{fig:end_matter} for each nucleus considered in this work. 
The \textit{ab initio} form factors from the dSCGF many-body method (blue) deviate markedly from the standard parameterization of \cite{Bjorken:2009mm, PhysRevD.8.3109} (black) for $Q \gtrsim 0.1$ GeV. This correction propagates directly through the cross section into the predicted DM signal yield. The two $\chi$EFT interactions considered in this work, N$^3$LO$_{\rm Texas}$ (blue, solid) and $\Delta$N$^2$LO$_{\rm GO}(394)$ (dark blue, dashed), yield similar form factors over the relevant momentum transfer range.

\paragraph{Event generator implementation.}
Both the elastic and QE channels are generated with \textsc{MadGraph5\_aMC@NLO} using a custom UFO model, in which the nuclear physics enters through externally tabulated structure functions applied as event level reweighting.

For elastic scattering, within \textsc{MadGraph},
the nucleus is modeled as a spin-0 state of mass $m_N$, the nucleus mass, coupling to the photon through a point-like vertex. This vertex is then dressed with the nuclear form factor by reweighting each event with the form factor $F(Q^2)^2$ from Eq.~\ref{eq:F(q)}. 
The nucleus photon vertex is given by,
\begin{equation}
    \Gamma^{el}_\mu = ieF(Q^2)\left( P_i + P_f\right)_\mu,
    \label{eq:elastic_vertex}
\end{equation}
where $P_i$ and $P_f$ are the initial and final state nucleus momenta.

For QE scattering, the beam electron instead scatters with a bound nucleon within the nucleus and in the final state the nucleon is unbound.
The target is represented by a free nucleon of mass $m_n$, the nucleon mass. The nuclear structure enters through the effective vertex,
\begin{equation}
    \Gamma^{QE}_\mu = G_1(|\mathbf{q}\,|,\omega)\,\gamma_\mu + \frac{G_2(|\mathbf{q}\,|,\omega)}{2m_n}\left(p_f+p_i\right)_\mu ,
\end{equation}
where $p_i$ and $p_f$ are initial and final state nucleon momenta, and with
\begin{equation}
\begin{aligned}
    G_1^2 &= \frac{W_1(|\mathbf{q}\,|,\omega)}{2m_n\omega} \\
    G_2^2 &= \frac{2m_nW_1(|\mathbf{q}\,|,\omega)/\omega - W_2(|\mathbf{q}\,|,\omega)}{1+\omega/2m_n},
\end{aligned}
\end{equation}
where the $G_1^2$ and $G_2^2$ contributions are generated as independent runs and summed at the cross section level, and $W_1$ and $W_2$ are computed from the spectral function described above.
The form of $\Gamma_\mu^{QE}$ follows from the impulse approximation, where the matrix element is approximated by the free nucleon electromagnetic current operator evaluated between the bound initial nucleon and the ejected, on-shell final nucleon, with the initial nucleon's momentum and removal energy distribution folded entirely into the spectral function. For an on-shell nucleon current, Lorentz invariance and current conservation (via the Gordon decomposition) restrict the vertex to two independent structures, $\gamma_\mu$ and $(p_f+p_i)_\mu$. The coefficients $G_1$, $G_2$ are fixed by matching the squared amplitude built from $\Gamma_\mu^{QE}$ (averaging over final state spins and summing over initial state spins) to the general, Lorentz covariant, and parity and gauge invariant nuclear tensor $W_{\mu\nu}$ (Eq. 5 of \cite{Benhar:2006wy}) decomposition in terms of $W_1(|\mathbf{q}\,|,\omega)$ and $W_2(|\mathbf{q}\,|,\omega)$.

\end{document}